\documentclass[3p, final]{elsarticle}

\usepackage{algorithm2e}
\usepackage{amssymb}
\usepackage{caption}
\usepackage{fancyvrb}
\usepackage{hyperref}
\usepackage{lineno}
\modulolinenumbers[5]

\usepackage{caption}
\usepackage{subcaption}
\usepackage{multirow}
\usepackage{scalerel}
\newcommand {\inlinefig}[1] {\scalerel*{\includegraphics{#1}}{|}}

\usepackage{tabularx}

\graphicspath{{image/}}

\nolinenumbers

\journal{Journal of Computers \& Security}

\begin{document}

\begin{frontmatter}

\title{Modeling and Analyzing Attacker Behavior in IoT Botnet using Temporal Convolution Network (TCN)\tnoteref{t1}\tnoteref{t2}}

\tnotetext[t1]{
    This research is supported by the National Science Foundation (NSF), USA, \href{https://www.nsf.gov/awardsearch/showAward?AWD_ID=1739032}{Award \#1739032.}
}

\tnotetext[t2]{
    The preliminary version of this paper was published as work in progress in IEEE International Conference on Communications 2021 (ICC 2021) \cite{sadique2021analysis}.
}

\author{Farhan Sadique}
\ead{fsadique@nevada.unr.edu}
\author{Shamik Sengupta}
\ead{ssengupta@unr.edu}
\address{Department of Computer Science and Engineering \\
University of Nevada, Reno, NV, USA}

\begin{abstract}

Traditional reactive approach of blacklisting botnets fails to adapt to the rapidly evolving landscape of cyberattacks. An automated and proactive approach to detect and block botnet hosts will immensely benefit the industry. Behavioral analysis of attackers is shown to be effective against a wide variety of attack types. Previous works, however, focus solely on anomalies in network traffic to detect bots and botnet. In this work we take a more robust approach of analyzing the heterogeneous events including network traffic, file download events, SSH logins and chain of commands input by attackers in a compromised host. We have deployed several honeypots to simulate Linux shells and allowed attackers access to the shells. We have collected a large dataset of heterogeneous threat events from the honeypots. We have then combined and modeled the heterogeneous threat data to analyze attacker behavior. Then we have used a deep learning architecture called a Temporal Convolutional Network (TCN) to do sequential and predictive analysis on the data. A prediction accuracy of $85-97\%$ validates our data model as well as our analysis methodology. In this work, we have also developed an automated mechanism to collect and analyze these data. For the automation we have used CYbersecurity information Exchange (CYBEX). Finally, we have compared TCN with Long Short-Term Memory (LSTM) and Gated Recurrent Unit (GRU) and have showed that TCN outperforms LSTM and GRU for the task at hand.

\end{abstract}


\end{frontmatter}

\nolinenumbers

\section{Introduction}\label{sec:intro}

A key component in many types of cyberattacks is a bot \cite{dunham2008malicious} -- a malicious program that allows an attacker to remotely control the infected host. Some notable examples of bot malware are Mirai \cite{antonakakis2017understanding}, Torpig \cite{stone2009your}, Conficker \cite{shin2010conficker} etc. A bot also connects the infected host to a botnet \cite{puri2003bots, feily2009survey} -- a network of such hosts. Botnets make up a large portion of the cybersecurity market.

Attackers use various techniques to infect a host with bot malware. For example, they can send the malware as an email attachment, post the download link on online forums and social networks, or host it on a website for drive-by downloads. They can also directly perform a brute-force attack to crack the password of a host.

Irrespective of how the host is infected, the attackers usually gains access to the `shell' of the compromised hosts. Therefore, they can use the botnet for a variety of cyberattacks such as adware distribution, DDoS attacks, hosting phishing websites, ransomware distribution, sending spam emails, spamming search engines, stealing credit card info etc. Thus, it is desirable for any person or any organization to detect and block botnet in their network.

The increasing adoption of the Internet of Things (IoT) have made IoT devices a major target of bots \cite{kolias2017ddos}. The most prominent example is the Mirai botnet \cite{antonakakis2017understanding} which compromises IoT devices using brute-force attack on the login credentials. It was first discovered in late 2016 and is still the most widespread botnet plaguing the IoT network.

IoT botnets, like Mirai, were able to take the internet by storm because of the proliferation of weakly configured IoT devices. A large number of IoT devices like refrigerators and CCTV cameras are configured with easily guessable usernames and passwords. Bot malware exploit that to perform a dictionary attack on the username/password pairs to gain access to the shell.

\subsection{Motivation}

The simplest defense against IoT botnets is manually blacklisting the IP addresses of the infected hosts. However, numerous hosts are compromised every day. At the same time, many compromised hosts become benign every day, as the owner regains control. Consequently, it is impossible to list all of their IPs. Moreover, blacklisting is a reactive approach because honeynets can . An IP shows up in a blacklist only after the host has done some harm. As a result, the industry would greatly benefit from a proactive defense mechanism against botnets. An intelligent system should detect a zero-day bot-host from its behavior not the IP. If a bot-host is detected in an early phase of the kill chain, it cannot do any harm to anybody.

Meanwhile, Intrusion detection systems (IDS) \cite{liao2013intrusion} use network signatures to detect bots. While they work really well for known patterns they cannot adapt to the new attacks. It also takes a long time to detect an attack pattern, analyze it and create its signature before adding it to an IDS \cite{karim2014botnet}. Another popular approach is using anomalies in network traffic \cite{karasaridis2007wide, binkley2006algorithm, gu2008botsniffer} to detect bots. Some works take it further to detect anomalies in DNS traffic \cite{choi2007botnet, villamarin2008identifying, dagon2005botnet, schonewille2006domain}.

The final approach is detecting anomalies in the infected hosts \cite{murugan2011system, creech2013semantic, ge2012effective}. The parameters used by these works are system calls, system API calls, syslogs, event logs etc. However, to the best of our knowledge no previous work has considered heterogeneous threat data to do behavior modeling. In particular, no previous work has considered correlating network traffic data, file download data and commands input into the shell to model attacker behavior. In addition, we have identified the following challenges in modeling attacker behavior in a botnet:

\begin{enumerate}
    \item The phrase `attacker behavior' is not well defined.
    \item There is no standard process or structure for modeling attacker behavior in a botnet.
    \item There is little automation in attacker behavior analysis, from data collection to data analysis.
    \item No previous work has considered multiple heterogeneous sources of threat data in modeling attacker behavior.
    \item Limited number of works have considered commands input into a compromised shell for modeling attackers.
\end{enumerate}

\subsection{Contribution}

This work considers heterogeneous threat for attacker behavior modeling, including network traffic, commands input into a compromised shell and files downloaded into the host. To the best of our knowledge no previous work has considered heterogeneous data for attacker behavior modeling. This is the novel contribution of this work. The contributions of this work are summarized below:

\begin{enumerate}
    \item We have collected a large dataset of heterogeneous threat data from bot infected hosts.
    \item We have clearly defined `attacker behavior' in this paper as a $4$ element vector.
    \item We have automated the whole process from data collection to analysis using CYbersecurity information Exchange (CYBEX) \cite{sadique2021cybersecurity, sadique2019system}.
    \item We have integrated multiple sources of threat data including network traffic, file downloads and shell commands.
    \item We have showed the efficacy of Temporal Convolutional Network (TCN) \cite{bai2018empirical} in predicting attacker behavior and compared it with Long Short-Term Memory (LSTM), and Gated Recurrent Unit (GRU).
    \item We have demonstrated the validity of our data model and TCN by predicting attacker behavior with an accuracy between $85-97\%$.
\end{enumerate}

\section{Related Work}\label{sec:rel}

Extensive research has been done on bot detection using anomalies in network traffic. Karasaridis et al. \cite{karasaridis2007wide} presented an algorithm to detect and characterize botnet by passive analysis of flow data. Their work is scalable and has a very low false positive rate. Binkley et al. \cite{binkley2006algorithm} presented an anomaly-based algorithm for detecting IRC-based botnet meshes. On the other hand, Gu et al. \cite{gu2008botsniffer} presented BotSniffer an approach that uses network-based anomaly to identity botnet command \& control channels. However, all these works deal with anomalies in network traffic data and do not present a robust methodology to model attacker behavior based on heterogeneous event types like we do in this work. Furthermore, none of these did any predictive analysis on attacker behavior.

Several works also considered anomalies in DNS traffic to detect bots and botnet. Choi et al. \cite{choi2007botnet} proposed a mechanism to detect botnet based on DNS traffic. Villamarín-Salomón et al. \cite{villamarin2008identifying} proposed another algorithm to detect bots based on their DNS requests. Dagon \cite{dagon2005botnet} presented yet another method to detect bots and botnets from DNS traffic. However, none of these works considered modeling the complete behavior of the attacker in bots based on various types of events. Moreover, in contrast to our work, none of these works did predictive analysis on the attacker behavior.

Shrivastava et al. \cite{shrivastava2019attack} analyzed commands input into the compromised shell of a Cowrie honeypot \cite{oosterhof2016cowrie} to classify different types of attacks. They have classified all the commands into $4$ categories -- malicious, DDOS, SSH and spying. They have also compared the accuracy of various classifiers including Naive Bayes, Random Forest and Support Vector Machine (SVM). However, our work considers features from not heterogeneous event types not just commands. Secondly, we predict the next move of the attacker to show the effectiveness of our model, rather than classifying them into several categories. Thirdly, they did not explain their feature collection methodology. Finally, our work is different from theirs because we do predictive analysis on attacker behavior which they did not do.

There are few previous works which did predictive analysis on attacker behavior.. Rade et al. \cite{rade2018temporal} modeled honeypot data using semisupervised Markov Chains and Hidden Markov Models (HMM). They also explored Long Short-Term Memory (LSTM) for attack sequence modeling. They concluded that LSTM provides better accuracy than HMM. However, they model Cowrie honeypot data as a $19$-state machine, where each state is defined by only one feature --`eventid' of Cowrie data. Cowrie `eventid's are explained in detail in subsections \ref{ss:cowrie_event} and \ref{ss:cowrie_event_types}. In our work we model the attacker data based on heterogeneous event types and define each state using $17$ features. We have also predicted $4$ different targets in this work in contrast to only $1$ target used in their work. One of the targets that we considered has $340+$ states compared to the $19$ states of their work. This makes our methodology more robust and versatile.

Deshmukh et al. \cite{deshmukh2019attacker} extended the work by Rade et al. \cite{rade2018temporal} to propose Fusion Hidden Markov Model (FHMM) for modeling attacker behavior. FHMM is more noise resistant and provides faster performance than  Deep  Recurrent  Neural  Network  (DeepRNN) with comparative accuracy in their analysis. However, since this work models the data based on the same $19$-state machine, it suffers from the same limitations as before.

In summary, there has been limited work done on modeling attacker behavior in a bot. Even less work has been done on predictive analysis of attacker behavior in the bots. In contrast to the previous works our work focuses on heterogeneous event types to model attacker behavior. This makes our work novel compared to the ones before.

\section{Background}\label{sec:back}

\subsection{Honeypot}

An honeypot is a decoy system. The Honeynet project \cite{spitzner2003honeynet} defines an honeypot as \textit{A security resource who’s value lies in being probed, attacked or compromised}. As honeypots have no production value, any activity logged in an honeypot can be deemed malicious.

Honeypots can be classified into two categories based on their purpose:

\begin{enumerate}
    \item Production honeypot: Production honeypots are used in campus networks to lure attackers away from production machines. They can also be used to detect attacker IP addresses or email addresses. They protect production servers by posing themselves easy targets for attackers.

    \item Research honeypot: Research honeypots are used to collect information. This information is further analyzed to detect new tools and techniques, to understand behavior of attackers and to detect attack patterns. Finally, the analyzed data can lead to newer defense techniques.
\end{enumerate}

Honeypots can be further classified into three categories on their level of interaction with the attacker:

\begin{enumerate}
    \item High Interaction Honeypots: Simulates all aspects of a real operating system (OS) completely. These honeypots can collect more information. However they are more risky to maintain, because the attacker can launch further attacks from these.
    \item Medium Interaction Honeypots: Simulates the aspects of an OS which cannot be used to launch further attacks.
    \item Low Interaction Honeypots: Simulates very basic aspects of an OS. They can collect very limited information and are low risk.
\end{enumerate}

\subsection{Cowrie Honeypot}

Cowrie \cite{oosterhof2016cowrie} is a medium to high interaction SSH and Telnet honeypot designed to log brute force attacks and the shell interaction performed by the attacker. In medium interaction mode (shell) it emulates a UNIX system in Python, in high interaction mode (proxy) it functions as an SSH and telnet proxy to observe attacker behavior to another system. In this work we use Cowrie in the medium interaction mode.

In our setup, cowrie only allows SSH logins into our honeypot. An attacker can login to the system using any username and password combination. Cowrie logs all the interactions including the source IP address of the attacker, the SSH parameters and the commands input while the attacker is logged in.

\subsection{CYBEX}

CYbersecurity information Exchange (CYBEX) \cite{sadique2019system, sadique2021cybersecurity} is a cybersecurity information sharing (CIS) platform with robust data governance. It automatically analyzes shared data to generate insightful reports and alerts. CYBEX-P has builtin software modules for data collection, data storage, data analysis and report generation. In this work we use CYBEX-P infrastructure as a service (IaaS) for collecting and analyzing the honeypot data.

\section{System Architecture}\label{sec:sysarch}

\begin{figure}[!ht]
    \includegraphics[width=.6\linewidth]{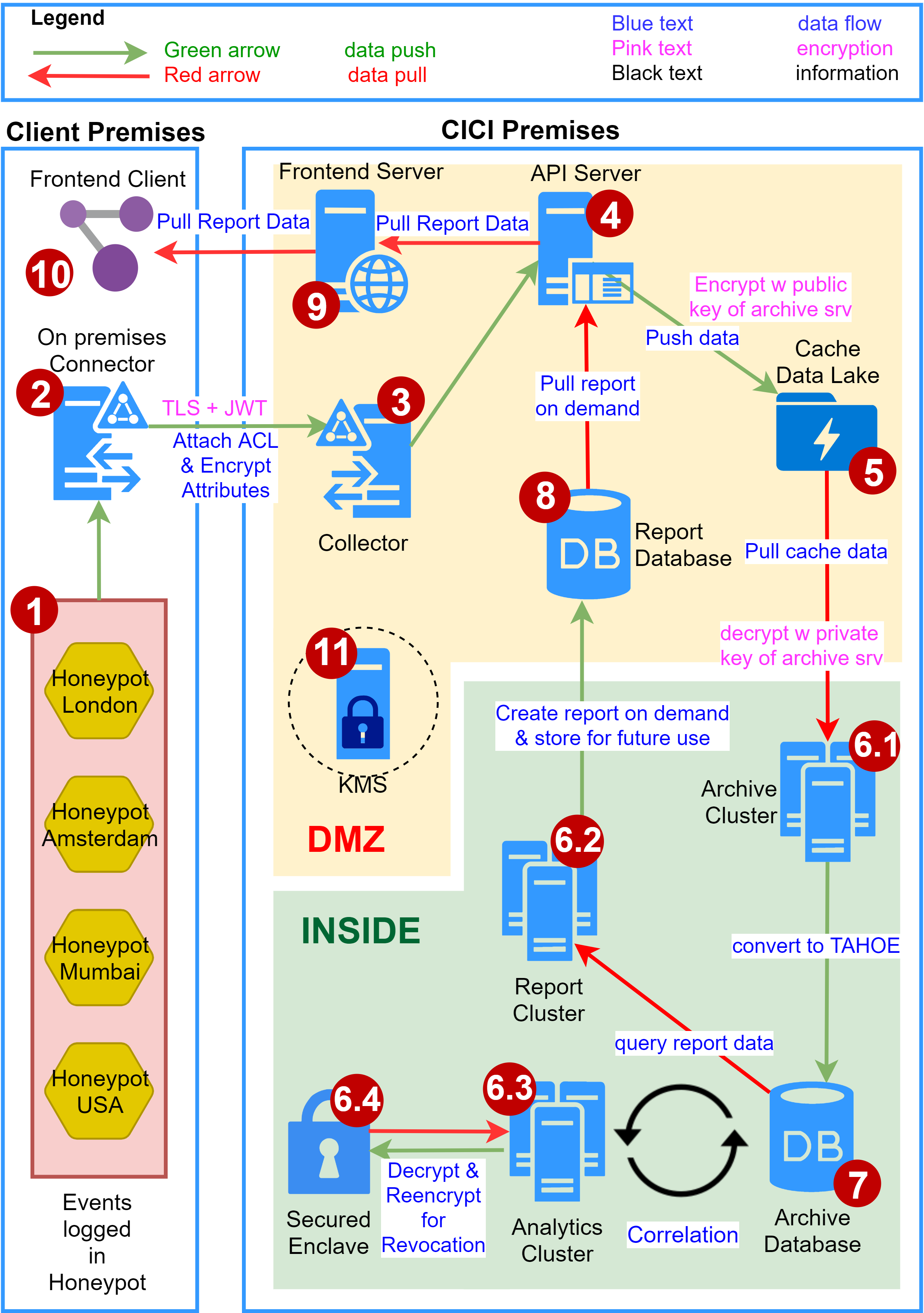}
    \centering
    \caption{System architecture of CYBEX along with the Data Flow.}
    \label{fig:sysarch}
\end{figure}

In this research, we have developed an automated framework to analyze and fingerprint attacker behavior in any compromised host. Our system uses CYbersecurity information Exchange (CYBEX) \cite{sadique2019system, sadique2021cybersecurity} infrastructure as a service (IaaS). CYBEX is a cloud based platform for organizations to share heterogeneous cyberthreat data. CYBEX accepts all kinds of human or machine generated data including firewall logs, emails, malware signatures etc. For this work we do not use the privacy module of CYBEX because all of our data are collected from publicly available honeypots. The related modules of CYBEX are described in this section. Our system has $7$ modules -- (1) Honeypots, (2) Frontend, (3) Input, (4) API, (5) Archive, (6) Analytics, and (7) Report. These modules share various components as shown in Fig. \ref{fig:sysarch}.

\subsubsection{Honeypots}

We have setup $5$ instances of the Cowrie honeypot all around the world. The locations are -- Amsterdam, Bangalore, London, Singapore and Toronto. All of them login SSH login attempts and corresponding commands input upon successful login. We have a diverse choice of four honeypots in four locations for better analysis and correlation.

\subsubsection{Frontend Module}\label{ss:fend}

The frontend module (\inlinefig{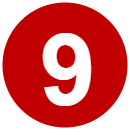}, \inlinefig{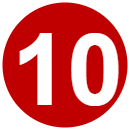} in Fig. \ref{fig:sysarch}) is a webapp for users to interact with CYBEX. This module allows users -- (1) to register with and login to CYBEX, (2) to configure the data sources, (3) to view the data, (4) to generate reports, and (5) to visualize the data.

\subsubsection{Input module}\label{subsec:input}

The input module (\inlinefig{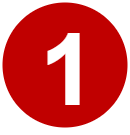}, \inlinefig{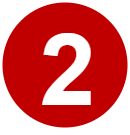}, \inlinefig{3}, \inlinefig{4}, \inlinefig{10} in Fig. \ref{fig:sysarch}) handles all the data incoming to CYBEX. Machine data is automatically sent via a connector (\inlinefig{2}) to the collector (\scalerel*{\includegraphics{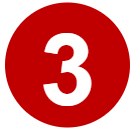}}{|}) using real time websockets. Afterwards, the collector posts the raw data to our API (\scalerel*{\includegraphics{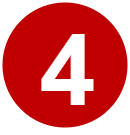}}{|}) endpoint. To ensure privacy, it uses the transport layer security (TLS) protocol \cite{dierks2008transport} during collection and posting.

\subsubsection{API module}\label{subsec:api}

The API module (\inlinefig{4}, \inlinefig{5} in Fig. \ref{fig:sysarch}) consists of the API server (\inlinefig{4}) and the cache data lake (\inlinefig{5}). It acts as the gateway for all data into and out of CYBEX. It serves two primary purposes:

\begin{enumerate}
	\item The input module (subsection \ref{subsec:input}) puts raw data into the system using the API.
	\item The report module (subsection \ref{subsec:report}) sends reports back to users using the API.
\end{enumerate}

\subsubsection{Archive module}\label{subsec:archive}

The archive module (\inlinefig{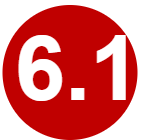}, \inlinefig{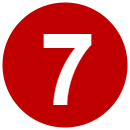} in Fig. \ref{fig:sysarch}) resides in the archive cluster and consists primarily of a set of parsing scripts. As mentioned earlier, the cache data lake (\inlinefig{5}) is encrypted with the public key of the archive server (\inlinefig{6_1}). The archive server -- (1) gets the encrypted data from the cache data lake (2) decrypts the data using own private key (3) parses the data into TAHOE, and (4) stores the data in the archive DB (\inlinefig{7}).

\subsubsection{Analytics module}\label{subsec:analytics}

The analytics module (\inlinefig{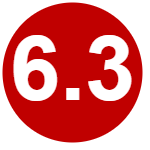}, \inlinefig{7} in Fig. \ref{fig:sysarch}) works on the archived data to transform, enrich, analyze or correlate them. It has various sub modules, some of which described here.

\paragraph{Filter sub-module}

An analytics filter parses a specific event from raw user data. Multiple filters can act on the same raw data and vice-versa.  For example, one filter can extract a \textit{file download event} from a piece of data while another filter can extract a \textit{DNS query event} from the same data.

\paragraph{Sequential Analysis sub-module}

This is a specialized sub-module that performs sequential analysis of the data based on the timestamp. It also correlates events in a session. A session is the time duration when one user is logged in.

\subsubsection{Report Module}\label{subsec:report}


Users use the report module (\inlinefig{4}, \inlinefig{5}, \inlinefig{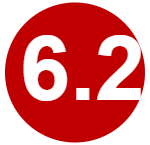}, \inlinefig{7}, \inlinefig{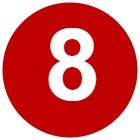}, \inlinefig{9}, \inlinefig{10} in Fig. \ref{fig:sysarch}) to generate and view reports. They request reports via the frontend client (\inlinefig{10},\inlinefig{9}). The API (\inlinefig{4}) stores the requests in the cache data lake (\inlinefig{5}). The report server (\inlinefig{6_2}) handles those requests by getting relevant data from the archive DB (\inlinefig{7}) and aggregating them into reports. It then stores the reports in the report DB (\inlinefig{8}). Users can access the reports on demand.

\section{Dataset}\label{sec:dataset}

\subsection{Data Source -- Cowrie Honeypot}

Cowrie \cite{oosterhof2016cowrie} is a medium to high interaction SSH and Telnet honeypot designed to log brute force attacks and the shell interaction performed by the attacker. In medium interaction mode (shell) it emulates a UNIX system in Python, in high interaction mode (proxy) it functions as an SSH and telnet proxy to observe attacker behavior to another system. In this work we use Cowrie in the medium interaction mode.

We have setup $5$ instances of Cowrie around the world -- Amsterdam, Banglaore, London, Singapore, Tornoto. The honeypots are configured to allow only SSH logins into the system. An attacker can login to the system using any username and password. Cowrie logs all the interactions including the source IP of the attacker, the login credentials, the SSH parameters, the downloaded files and the commands input into the shell.

\subsection{Cowrie Data as Events}\label{ss:cowrie_event}

\begin{figure}[!htbp]
    \centering
    \footnotesize
    \begin{BVerbatim}
{
  "eventid": "cowrie.session.file_download",
  "timestamp": "2020-04-28T00:00:22.134604Z",
  "src_ip": "5.188.87.49",
  "session": "d151a9c7",
  "sensor": "london",
  ...
}
    \end{BVerbatim}
    \normalsize
    \caption{Common attributes of all Cowrie events.}
    \label{fig:cowrie}
\end{figure}

Cowrie structures collected data into events. Fig. \ref{fig:cowrie} shows the common attributes of a Cowrie event. These attributes are explained below --

\begin{enumerate}
    \item \textit{eventid:} Denotes the type of the event. The `eventid' \textit{cowrie.session.file\_download} in Fig. \ref{fig:cowrie} means, it was generated when the attacker downloaded a file into the compromised machine.
    \item \textit{timestamp:} The time when the event was recorded by the honeypot.
    \item \textit{src\_ip:} Source IP address of the attacker.
    \item \textit{session:} A session is a sequence of events generated during one login session. Cowrie maintains a unique `session' for each session and assigns the same session-ID to the events of a session.
    \item \textit{sensor:} Hostname of the honeypot server.
\end{enumerate}

\subsection{Cowrie Event Types}\label{ss:cowrie_event_types}

Cowrie generates about $20$ different `eventid's. Many of these event types are related to the SSH session, key exchange and logging and do not carry valuable information. For this work, we are interested in the following `eventid's --

\begin{itemize}
    \item \textit{cowrie.login}: Generated when the attacker tries to SSH into the host. Contains the username and the password of the SSH request. Fig. \ref{fig:login.success} shows a \textit{cowrie.login} event. Here the suffix \textit{.success} means that the login attempt was successful. Note that the common attributes shown in Fig. \ref{fig:cowrie} are omitted from Fig. \ref{fig:login.success}.

\begin{figure}[!ht]
    \centering
    \footnotesize
    \begin{BVerbatim}
{
  "eventid": "cowrie.login.success",
  "password": "asdasdasd",
  "username": "root",
  ...
}
    \end{BVerbatim}
    \normalsize
    \caption{Attributes of \texttt{cowrie.login} event.}
    \label{fig:login.success}
\end{figure}

    \item \textit{cowrie.direct-tcpip}: Generated when the attacker tries to communicate over the internet through the TCP-IP protocol. Contains the destination IP, destination port and the data (if present). The suffic \textit{.data} means that this communication contains data. Fig. \ref{fig:direct.tcp-ip} shows a \textit{cowrie.direct.tcp-ip} event. Note that the `data' field is truncated. Also, the common attributes from Fig. \ref{fig:cowrie} are omitted.

\begin{figure}[!ht]
    \centering
    \footnotesize
    \begin{BVerbatim}
{
  "data": "\\x03\\x00\\xa6...",
  "dst_ip": "www.walmart.com",
  "dst_port": 443,
  "eventid": "cowrie.direct-tcpip.data",
  ...
}
    \end{BVerbatim}
    \normalsize
    \caption{Attributes of \texttt{cowrie.direct.tcp-ip} event.}
    \label{fig:direct.tcp-ip}

\end{figure}

    \item \textit{cowrie.session.file\_download}: Generated when the attacker downloads a file into the compromised host. Contains the download URL, the file hash and the actual binary of the file. Fig. \ref{fig:session.file-download} shows a \textit{cowrie.session.file\_download} event.

\begin{figure}[!htbp]
    \centering
    \footnotesize
    \begin{BVerbatim}
{
  "eventid": "cowrie.session.file_download",
  "outfile": "dl/6e223babfbd3e...",
  "shasum": "6e223babfbd3eef8...",
  "url": "http://192.210.236.38/bins.sh"
  ...
}
    \end{BVerbatim}
    \normalsize
    \caption{Attributes of \texttt{cowrie.session.file\_download} event.}
    \label{fig:session.file-download}

\end{figure}

    \item \textit{cowrie.command}: Generated when the attacker inputs a command into the shell of the compromised host. It contains the exact command. The suffix \textit{.success} means the command was simulated successfully. Fig. \ref{fig:command.success} shows a \textit{cowrie.command} event. The \textit{input} attribute contains the exact command input by the attacker in the shell.

\begin{figure}[!htbp]
    \centering
    \footnotesize
    \begin{BVerbatim}
{
  "eventid": "cowrie.command.success",
  "input": "cat /proc/cpuinfo",
  ...
}
    \end{BVerbatim}
    \normalsize
    \caption{Attributes of \texttt{cowrie.command} event.}
    \label{fig:command.success}
\end{figure}

\end{itemize}

\subsection{Command, Parameter \& Type}\label{ss:cmd_type}

After logging in, the attackers often execute different commands in the honeypot shell. These commands are documented by Cowrie in the \textit{cowrie.command} events. An example command is \texttt{wget NasaPaul.com/v.py}. Here \texttt{wget} is the actual command and \texttt{NasaPaul.com/v.py} is its parameter.

In our database we have seen about $40$ unique commands. However, many of these commands, like \texttt{wget} take parameters and there are more than $300$ unique command-parameter combinations. We have further classified these commands into $7$ types based on the intention of the attacker. These types are:


\begin{enumerate}
    \item System Info – Check software, hardware or system configuration.

    \item Cover Track – Hide evidence of intrusion and malicious activity.

    \item Install – Install a software in the system that was not previously in there.

    \item Download – Download a remote file into the honeypot system.

    \item Run – Run or execute a program or a script.

    \item Escalate privilege – Change password or gain root access to the system.

    \item Change config – Change system configuration including hostname, network and firewall configuration.

\end{enumerate}

\begin{figure}[!ht]
    \includegraphics[width=0.9\linewidth]{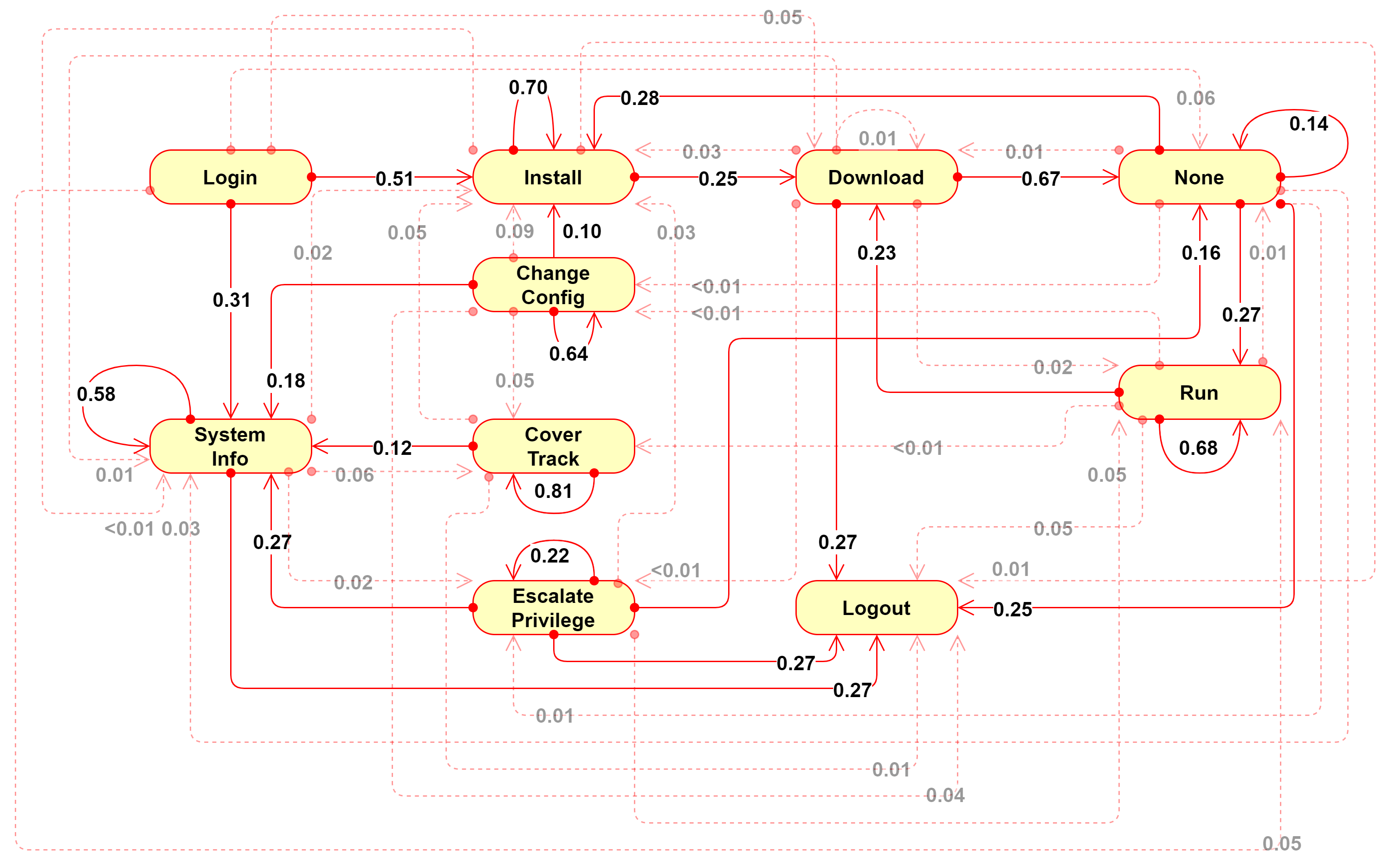}
    \centering
    \caption{Representation of Cowrie sessions as a finite state machine of the command types.}
    \label{fig:state}
\end{figure}

\begin{figure}[!ht]
\includegraphics[height=.7\textheight]{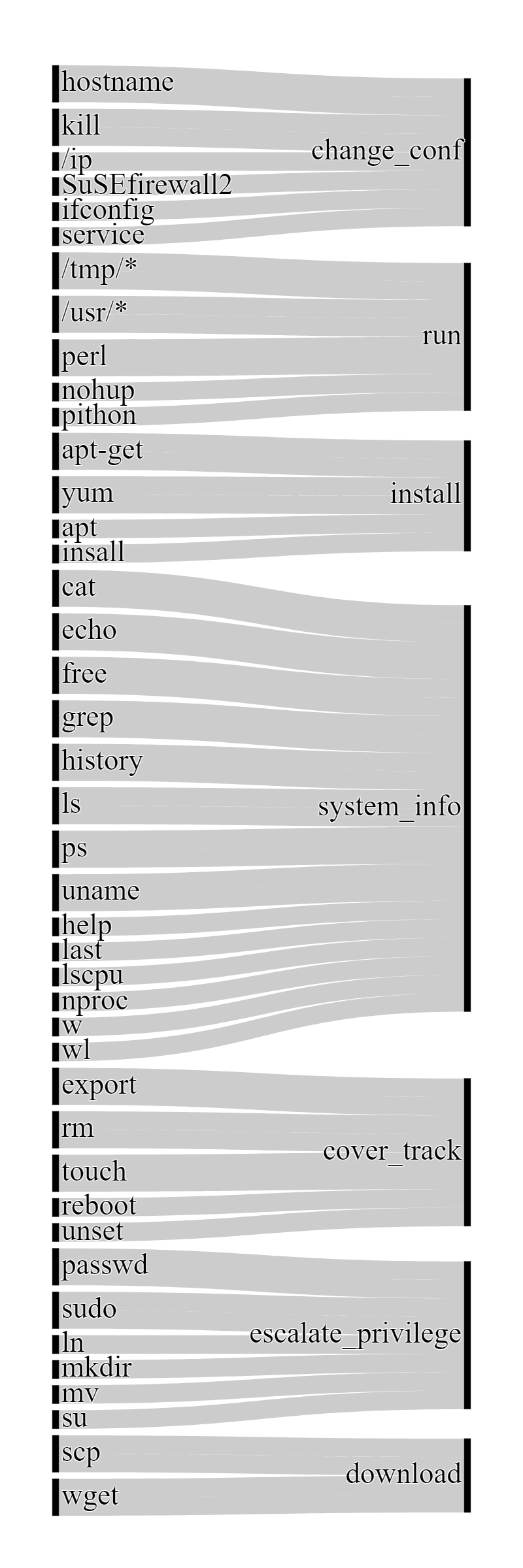}
\centering
\caption{Map of commands to command types.}
\label{fig:command_type}
\end{figure}

Fig. \ref{fig:state} shows the Cowrie sessions as a state machine of these command types. The edges are the probabilities of transition from one state to the next. The state \texttt{None} means the event is not of type shell command. Table \ref{tbl:command} shows the different command types and corresponding commands. The classification of commands into command types is further visualized in Fig. \ref{fig:command_type}.

\begin{table}[!htbp]
    \caption{Classification of commands into different types.}
    \label{tbl:command}
    \centering
    \begin{tabularx}{\linewidth}{|c|X|}
        \hline
        \textbf{Command Type} & \textbf{Commands} \\

        \hline
        System Info & cat, echo, free, help, history, last, ls, ps, w, grep, lscpu, nproc, uname, wl \\
        \hline
        Cover track & export, reboot, rm, touch, unset \\
        \hline
        Install & apt, apt-get, install, yum \\
        \hline
        Download & scp, wget \\
        \hline
        Run & nohup, perl, python, /tmp/*, /usr/* \\
        \hline
        Escalate privilege & ln, mkdir, mv, passwd, su, sudo \\
        \hline
        Change config & hostname, ifconfig, /ip, kill, susefirewall2, service \\
        \hline
    \end{tabularx}
\end{table}

\subsection{Dataset at a Glance}\label{ss:datastat}

\begin{enumerate}
    \item Total number of events = $3567$
    \item Total number of sessions (sequences) = $393$
    \item From date = April 4, 2020
    \item To date = May 8, 2020
    \item Number of Cowrie honeypots = 5
    \item Location of honeypots = Amsterdam, Bangalore, London, Singapore, Toronto
\end{enumerate}

\section{Data Processing}\label{sec:dataproc}

As discussed in section \ref{sec:dataset}, cowrie generates heterogeneous data with different attributes. For example, a \textit{cowrie.direct-tcpip} event has the \textit{dst\_ip} and \textit{dst\_port} attributes, which other event types do not have. In this section we extract features from such heterogeneous and group them together into a single feature table so that they can be fed into a learning model.

As described in section \ref{sec:sysarch}, We have used CYBEX to automate the entire procedure of data collection to data analysis for this work. In other words we have used CYBEX infrastructure as a service (IaaS) for this work. This work is closely coupled with the development of CYBEX.

Along with CYBEX, we have further developed TAHOE a graph-based cyberthreat language (CTL). In this section, we discuss the modeling of Cowrie data in TAHOE format. TAHOE offers several advantages over traditional CTLs -- Firstly, TAHOE can store all types of structured data. Secondly, queries in TAHOE format are faster than in other CTLs. Thirdly, TAHOE intrinsically correlates the heterogeneous data. Finally, TAHOE is scalable for all kinds of data analysis - a major limitation of other CTLs.

In this section, we further discuss how to featurize such heterogeneous data to use them for machine learning. We start by explaining the complete lifecycle of the data in CYBEX from data generation.

\subsection{Data Flow in CYBEX}\label{ss:data_flow}

\subsubsection{Data Generation}

Each Cowrie honeypot (\inlinefig{2} in Fig. \ref{fig:sysarch}) simulates a generic IoT device. They generate data in the format of Fig. \ref{fig:cowrie}. The honeypots log these data pieces into a file in respective server. We call each such log message a raw document.

\subsubsection{Data Input}

Each of our honeypot installations have a connector agent (\inlinefig{2} in Fig. \ref{fig:sysarch}). The connector is basically a script that reads the raw data from log files and sends them to the CYBEX collector (\inlinefig{3}) via a realtime websocket. The data in transport are encrypted via TLS.

\subsubsection{Data Collection}\label{ss:datacoll}

The collector then posts the data to the API (\inlinefig{4}). The API encrypts the data with the public key of the archive cluster (\inlinefig{6_1}) and stores the encrypted data in the cache data lake (\inlinefig{5}).

\subsubsection{Data Archiving}

\begin{figure}[!htbp]
    \centering
    \footnotesize
    \begin{BVerbatim}
{
  "itype": "raw",
  "data": {
    "eventid": "cowrie.session.file_download",
    "timestamp": "2020-04-28T00:00:22.134604Z",
    "src_ip": "5.188.87.49",
    "session": "d151a9c7",
    "sensor": "london",
    ...
  },
  "sub_type": "cowrie_honeypot",
  "timezone": "US/Pacific",
  "_hash": "3d5792b...",
  ...
}
    \end{BVerbatim}
    \normalsize
    \caption{A Cowrie event encapsulated in a TAHOE \texttt{raw} document.}
    \label{fig:raw}
\end{figure}

The archive cluster (\inlinefig{6_1}), then pulls the data from the cache data lake (\scalerel*{\includegraphics{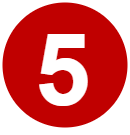}}{|}), decrypts the data using its private key, converts the cowrie events into TAHOE \texttt{raw} format and stores them in the archive database (\inlinefig{7}). TAHOE \texttt{raw} basically puts a wrapper around the Cowrie event.

Fig. \ref{fig:raw} shows a TAHOE \texttt{raw} document. The \texttt{\_hash} (truncated in figure) is the unique ID of the document and generated as the SHA256 checksum of the \texttt{data} field. CYBEX collects different types of data; so TAHOE uses the \texttt{sub\_type} field to distinguish between them.

\subsubsection{Data Filtering}\label{sss:data_filtering}

Data filtering in CYBEX means parsing a TAHOE \texttt{raw} document into TAHOE \texttt{events}. Each \texttt{sub\_type} of a TAHOE \texttt{raw} document represents a different type of data; and has its own \textit{filter} scripts. The \textit{filter} basically extracts different attributes from the \texttt{raw} document and restructures them into TAHOE \texttt{events}.

\begin{figure}[!htbp]
    \centering
    \footnotesize
    \begin{BVerbatim}
{
  "itype" : "event",
  "timestamp" : 1588093536.969,
  "category" : "unknown",
  "data" : {
    "success" : [false],
    "shell_command" : ["cat /proc/cpuinfo"],
    "attacker" : [{
        "ipv4" : ["134.122.20.113"]
    }]
  },
  "_cref" : [
    "e7dc7351c504da69f7a43421...,
    "966fca3ed576e47e9d2ae2a7...,
    "a58a2e656c004f01b38dc77c...
  ],
  "sub_type" : "shell_command",
  "_hash" : "b3da61a6313307f739...",
  ...
}
    \end{BVerbatim}
    \normalsize
    \caption{A TAHOE \texttt{event} document.}
    \label{fig:event}
\end{figure}

This parsing is done by the analytics cluster (\inlinefig{6_3}). It reads the \texttt{raw} data from the archive database (\inlinefig{7}), parses the data and writes the results back in the archive database. Fig. \ref{fig:event} shows the structure of a TAHOE \texttt{event}.

The \texttt{sub\_type} of a TAHOE event depends on the \textit{eventid} of the original \texttt{raw} document. For example a \textit{cowrie.login} event is parsed into a TAHOE \texttt{ssh} event. The mapping is -- \textit{cowrie.command} $\Rightarrow$ \texttt{shell\_command}, \textit{cowrie.direct.tcp-ip} $\Rightarrow$ \texttt{network\_traffic}, \textit{cowrie.session.file-download} $\Rightarrow$ \texttt{file\_download}, \textit{cowrie.login} $\Rightarrow$ \texttt{ssh}. So, TAHOE basically normalizes the Cowrie events into the standardized format - TAHOE.

Notice that, the Cowrie \textit{session} ID is not stored in TAHOE \texttt{events}. For that, we use a separate TAHOE structure called a \texttt{session}. Fig. \ref{fig:session} shows the structure of a TAHOE \texttt{session}. The field \texttt{\_ref} stores the ID of all the \texttt{events} that belong to this session. So, basically it forms a directed graph with the \texttt{session} node as the root and the \texttt{events} as leaves.

This concludes the default flow of any threat data through CYBEX. We have now converted Cowrie events into TAHOE events without any loss of information. Next, we begin processing the data for this particular task of attacker behavior modeling.

\begin{figure}[!htbp]
    \centering
    \footnotesize
    \begin{BVerbatim}
{
  "itype" : "session",
  "data" : {
    "hostname" : ["london"],
    "sessionid" : ["5a0facf9"]
  },
  "_cref" : [
    "53df245bcefb3f2a558349c37...",
    "39885eec34b95fa2acdfffd14..."
  ],                            ..."
  "_ref" : [                    ..."
    "b3da61a6313307f7394510146...",
    "1e31784145b52a42d964f5a5c...",
    "cb3be781b29297571cc20cbf8...",
      ...                       ..."
  ],                            ..."
  "sub_type" : "cowrie_session",..."
  "_hash" : "98601c106789882a4ee...",
  "start_time" : 1588924147.615,
  "duration" : 3.29502511024475,
  "end_time" : 1588924147.616
}
    \end{BVerbatim}
    \normalsize
    \caption{A TAHOE \texttt{session} document.}
    \label{fig:session}
\end{figure}

\subsection{Advantages of using CYBEX \& TAHOE}

The data collected from Cowrie honeypot is used as an example to validate our methodology in this work. However, we want to propose this methodology for all types of threat data collected from heterogeneous sources. This is particularly problematic because different sources store or log data in their own format. For example two firewalls from two different vendors will collect network traffic logs in different formats. CYBEX automatically recognizes the sources and normalizes those seemingly different data into TAHOE. TAHOE acts as the standardized format here while CYBEX acts as the automated parser.

Each event has a different set of attributes or properties. This makes them unsuitable for storing as a row in a relational database. TAHOE uses a JSON structure to store such arbitrarily structured events. Morever, there could be any number of edges or connections between the attributes, objects, events and sessions. Such arbitrary length of the edges array is again unsuitable to be stored in a cell in a relational database. However, the JSON structure poses no such limitation on TAHOE. Finally, TAHOE differs itself from other JSON based CTI (e.g. STIX) by being indexable. As a result, we can query events connected to an attribute or a group of events connected to a session really fast. To the best of our knowledge, there is no such CTI available in the industry right now.

\subsection{Sequence of Events}

In subsection \ref{ss:data_flow} we saw how CYBEX parses any threat data into TAHOE events. Now, we further curate these TAHOE events for the task at hand - attacker behavior modeling. In this work we do not consider each event independent. Rather we are interested in modeling attacker behavior as a sequence of events.

\begin{figure}[!htbp]
    \includegraphics[height=1.62in]{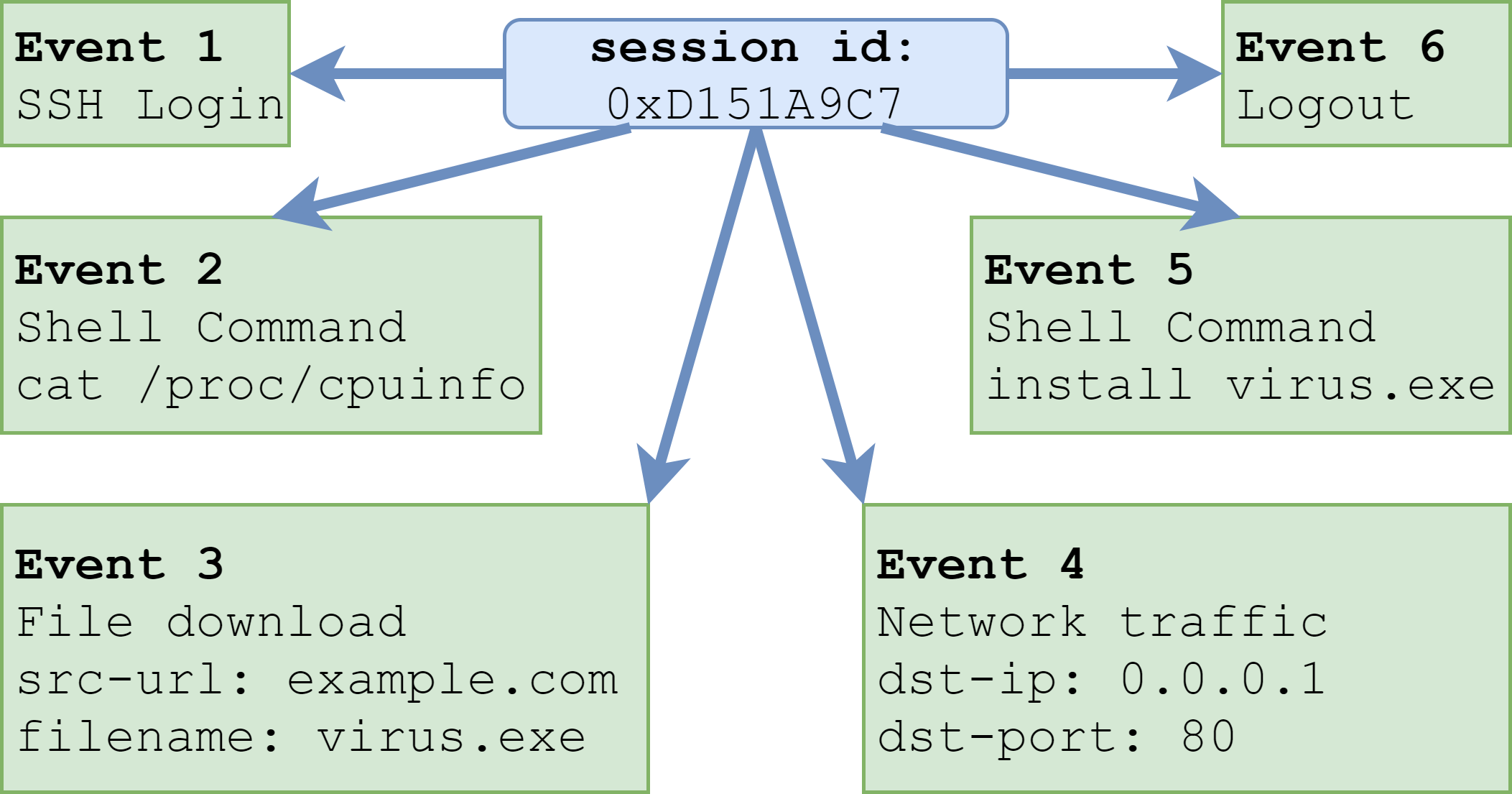}
    \centering
    \caption{A TAHOE \texttt{session} with $6$ \texttt{events} as a directed graph.}
    \label{fig:session_graph}
\end{figure}

As stated in subsection \ref{ss:cowrie_event}, Cowrie generates a unique \textit{session} ID whenever an attacker logs into the honeypot. Cowrie stores this ID in all events generated during this login session. We can use that ID to group the events in a session. We can then sort these events by their timestamps to form a sequence.

Fig. \ref{fig:session_graph} shows an example sequence of events. Here the attacker logs into the honeypot, executes a command in the shell, downloads a file, sends some data over TCP/IP, executes another command and then logs out. There are $6$ events in this example session. However, there can be any number of events in a session. In our dataset we have seen a minimum of $2$ events and a maximum of $47$ events in some sessions.

At this point, we group all events into event-sequences like that of Fig. \ref{fig:session_graph}. We then end up with a number of sequences, each with an arbitrary number of events. We then move onto extracting features for each of these events.

\subsection{Feature Extraction}\label{ss:featex}

Now, for each event in a session we extract the following features:

\newcounter{numFeature}

\paragraph{Time related}
\begin{enumerate}
    \item Hour (integer): Hour of day.
    \item Date (integer): Date of month.
    \item Month (categorical): January, February etc.
    \item Day (categorical): Sunday, Monday etc.

    \setcounter{numFeature}{\value{enumi}}
\end{enumerate}

\paragraph{Session Related}
\begin{enumerate}
    \setcounter{enumi}{\value{numFeature}}

    \item Session start (boolean): Is this event at the beginning of a session?
    \item Session end (boolean): Is this event at the end of a session?
    \item Event order (integer): How many events have been recorded in this session?

    \setcounter{numFeature}{\value{enumi}}
\end{enumerate}

\paragraph{Other common features}
\begin{enumerate}
    \setcounter{enumi}{\value{numFeature}}

    \item Event type (categorical): Valid event types are ssh, network traffic, shell command and file download as described in subsubsection \ref{sss:data_filtering}.
    \item Sensor (categorical): Location of the honeypot server - Amsterdam, Bangalore, London, Singapore or Toronto.
    \item Attacker IP (categorical): IP address of the attacker.

    \setcounter{numFeature}{\value{enumi}}
\end{enumerate}

\paragraph{Only for \texttt{cowrie.direct.tcp-ip} events}
\begin{enumerate}
    \setcounter{enumi}{\value{numFeature}}

    \item Destination IP (categorical): Destination IP address of the TCP-IP packet/s.
    \item Destination Port (integer): Destination port of the TCP-IP packet/s.

    \setcounter{numFeature}{\value{enumi}}
\end{enumerate}

\paragraph{Only for \texttt{cowrie.command} events}
\begin{enumerate}
    \setcounter{enumi}{\value{numFeature}}

    \item Command + parameter (categorical): The shell command with parameter.
    \item Command (categorical): This feature is derived from `command + parameter' and lists the actual shell command without parameter.
    \item Command type (categorical): This is another derived feature and lists the type of command as described in subsection \ref{ss:cmd_type}.
    \item Command success (boolean): Was the command successfully simulated?

    \setcounter{numFeature}{\value{enumi}}
\end{enumerate}

\paragraph{Only for \texttt{cowrie.login} events}:
\begin{enumerate}
    \setcounter{enumi}{\value{numFeature}}

    \item Login success (boolean): Was the attacker successful to login?
\end{enumerate}

For example, if we extract the features of the event in Fig. \ref{fig:event}, we get the feature vector shown in Fig. \ref{fig:feat_vec}. Note that, it does not have any valid value for the features `Dest IP' and `Dest Port', because these two features are defined for `network traffic' events only. Similarly, `login success' is defined for `ssh' events only. Also note that, the `session start', `session end', `event order' and `sensor' features are not directly extracted from the event data in Fig. \ref{fig:event}. Rather these are extracted from the session data in Fig. \ref{fig:session}. We can look this up in our database, because the session in Fig. \ref{fig:session} contains the \texttt{\_hash} of this event in its \texttt{\_ref} field.

\begin{figure}[!htbp]
    \centering
    \footnotesize
    \begin{BVerbatim}
Common features:
================

Hour  Date  Month  Day  Session  Sesssion  Event    Event  Sensor        Attacker
                          Start       End  Order     Type                      IP
----  ----  -----  ---  -------  --------  -----  -------  ------  --------------
  17    28    Apr  Tue     True     False      1    Shell  London  134.122.20.113
                                                  Command


Features realated to particular event type:
===========================================

Dest  Dest          Command +  Command  Command  Command    Login
  IP  Port          Parameter              Type  Success  Success
----  ----  -----------------  -------  -------  -------  -------
None  None  cat /proc/cpuinfo      cat   System    False     None
                                           Info
    \end{BVerbatim}
    \normalsize
    \caption{Feature vector of the event in Fig. \ref{fig:event}.}
    \label{fig:feat_vec}
\end{figure}

At this point we have a number of sequences like that of Fig. \ref{fig:session_graph}. Each sequence has arbitrary number of events and each of those events are represented by a $17$ parameter vector like Fig. \ref{fig:feat_vec}. With this dataset we are ready to define the problem statement of our analysis methodology.

\section{Analysis Methodology}\label{sec:meth}

So far we have modeled attacker behavior as a sequence of events. We have also represented each of those events as a $17$ element vector. In this section we asses the validity of our model with real data. We do this by predicting future attacker behavior based on past events. If we can successfully predict the next step an attacker takes, we can simultaneously infer that attacker behavior is predictable and our model is valid.

To show this, we have chosen to predict the following targets -- (1) event type ($4$ different values), (2) shell command with parameter ($300+$ different values), (3) shell command ($40+$ different values) and (4) shell command type ($7$ different values). We call this set of targets the `attacker behavior' for a particular event.

This is a sequence modeling problem, because an event in a sequence depends on the previous events. Recurrent neural networks (RNN), like Long Short-Term Memory (LSTM) and Gated Recurrent Unit (GRU), are better suited for sequence modeling \cite{goodfellow2016deep}.  However, a recent publication \cite{bai2018empirical} has showed viability of a convolutional neural network called a temporal convolutional network (TCN) as well. So, in this section we compare three neural networks to predict attacker behavior -- (1) TCN, (2) LSTM, and (3) GRU.

\subsection{Problem Statement}

Here, the set of predictors is $\mathbb{P} = $ \{hour, date, month, day, session start, session end, event order, event type, sensor, attacker ip, dest ip, dest port, command + parameter, command, command type, command success, login success\}. And, the set of targets is $\mathbb{T} = $ \{event type, command + parameter, command, command type\}

Now, let us assume, $\mathbb{S} = {S_1, S_2, ..., S_N}$ is a set of $N$ sequences. Here, $S_i = <E_1, E_2, ... \allowbreak, E_{M_i}>$ is the $i^{th}$ sequence with $M_i$ events in it.

Then, $\mathbb{P}_{i,j} = <x1_{i,j}, x2_{i,j}, ..., x17_{i,j}>$ is the vector of the $17$ features in $\mathbb{P}$ for the $j^{th}$ event in the $i^{th}$ sequence. This vector contains our predictors.

Also, the target vector $\mathbb{T}_{i,j} = <z1_{i,j}, z2_{i,j}, ..., z4_{i,j}>$ is the vector of the $4$ targets in $\mathbb{T}$ for the $j^{th}$ event in the $i^{th}$ sequence. $\mathbb{T}_{i,j}$ represents attacker behavior in our work.

We want to find the function $f$ which minimizes some expected loss between the targets and the predicted values $\mathcal{L}( \mathbb{T}, f(\mathbb{P}))$.

\subsection{Temporal Convolutional Network (TCN)}

\begin{figure}[!ht]
    \includegraphics[width=0.7\linewidth]{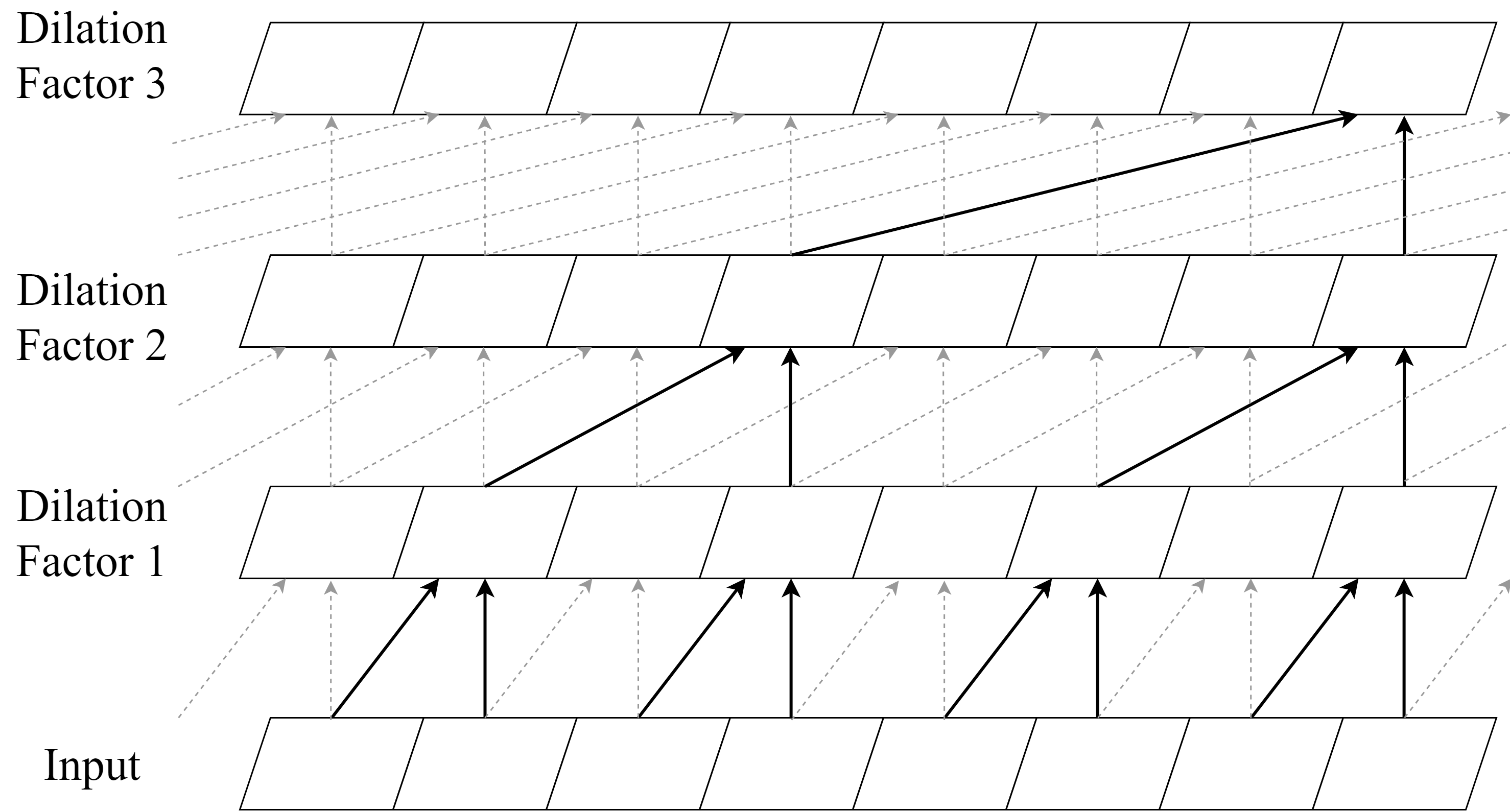}
    \centering
    \caption{Dilated causal convolutional layers of a typical TCN.}
    \label{fig:tcn1}
\end{figure}

Convolutional neural networks (CNNs) are commonly associated with image classification tasks. However, Bai et al. \cite{bai2018empirical} outlined the general structure for a temporal convolutional networks (TCN) which can be used to create a robust prediction model for sequences. They have also empirically showed how TCN matches or even outperforms traditional recurrent neural networks (RNNs) in sequence modeling and prediction.

The principal building block of TCN is a dilated causal convolution layer. Here, `causal' means the output for the current step do not depend on future steps. Also dilated convolutions are used to increase the receptive field of the layers. Multiple such layers can be stacked to form a deeper network. The dilation factor is increased exponentially as shown in Fig. \ref{fig:tcn1}.

\begin{figure}[!ht]
    \includegraphics[width=0.5\linewidth]{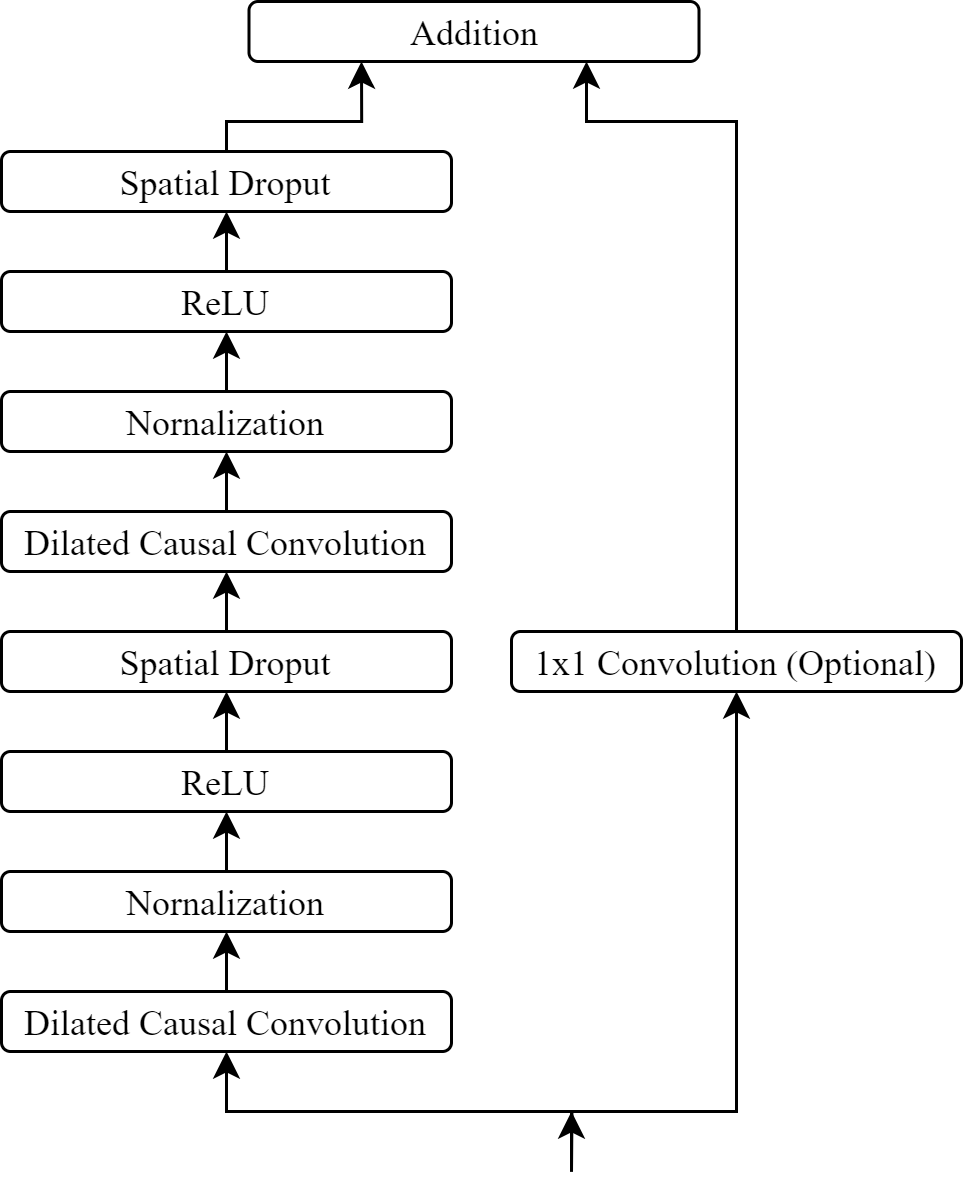}
    \centering
    \caption{A TCN Residual Block.}
    \label{fig:tcn2}
\end{figure}

The architecture of a general TCN described in  \cite{bai2018empirical} contains multiple residual blocks. Each residual block consists of two dilation causal convolution layers with same dilation factor along with normalization, ReLU activation and dropout layers. The input to each residual block is also added to the output when the number of channels between the input and the output are different. A general residual block is shown in Fig. \ref{fig:tcn2}.

\begin{figure}[!ht]
    \includegraphics[width=.7\linewidth]{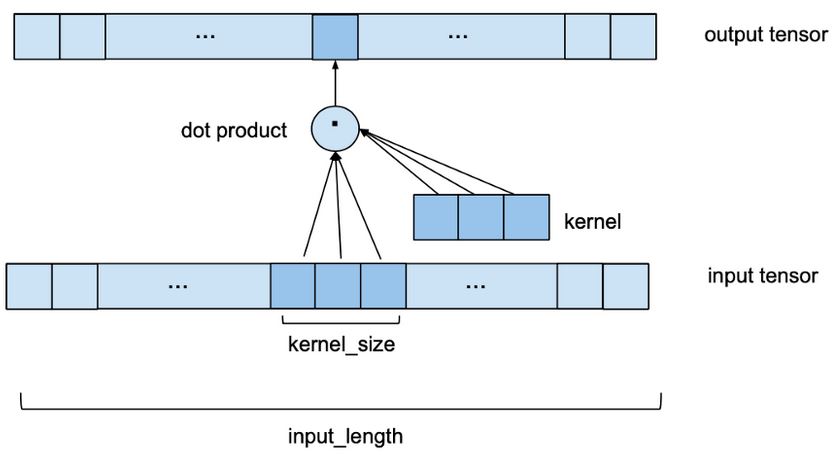}
    \centering
    \caption{General architecture of a TCN Classifier.}
    \label{fig:tcn}
\end{figure}

We can then put one or more such residual blocks into a general sequence classifier to get the architecture of a TCN classifier as shown in Fig. \ref{fig:tcn}. The network begins with a sequence input layer followed by one or more residual blocks. The residual blocks are then followed by a fully connected layer, a softmax layer, and a classification output layer.

\subsection{Long Short-Term Memory (LSTM)}

LSTM \cite{hochreiter1997long} is an artificial recurrent neural network (RNN) architecture \cite{mikolov2011extensions} used by deep learning practitioners for sequence modeling. LSTM has feedback connections in contrast to traditional feed forward neural networks. As a result, LSTMs can learn long-term dependencies. This property makes LSTM suitable for sequence modeling and predictions.

\begin{figure}[!ht]
    \includegraphics[width=.4\linewidth]{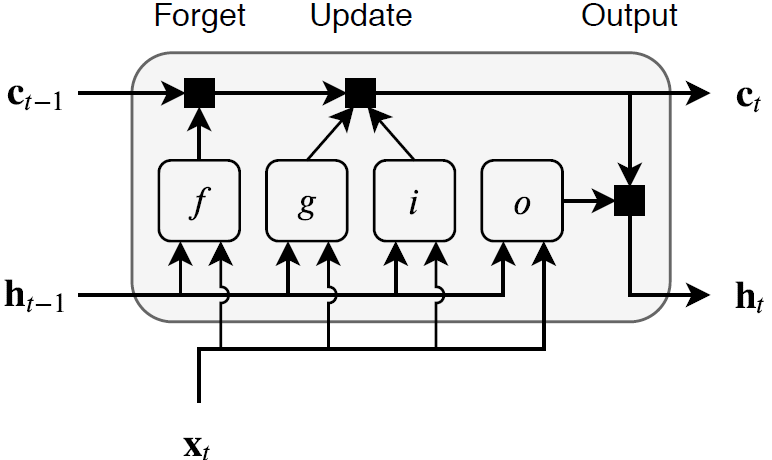}
    \centering
    \caption{An LSTM Block.}
    \label{fig:lstm1}
\end{figure}

The core component of an LSTM network is an LSTM block as shown in Fig. \ref{fig:lstm1}. Here, $c_t$ is the cell state at time-step (sequence step) $t$ whereas $h_t$ is the hidden state also called the cell output. The forget gate, $f$, determines which values to remove from the cell state, whereas the input gate, $i$, controls which values to update. The actual update-values are determined by the memory gate, $g$. Finally, the output gate, $o$ controls which values to output.

\begin{figure}[!htbp]
    \includegraphics[width=.8\linewidth]{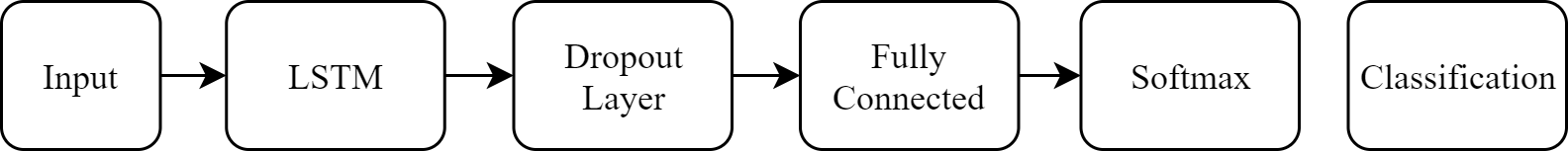}
    \centering
    \caption{A general LSTM Classifier.}
    \label{fig:lstm}
\end{figure}

Each element of a sequence passes through the LSTM block and updates it, forming an LSTM layer. Just like TCN we can place this LSTM layer inside a general sequence classifier to get the architecture of an LSTM classifier as shown in Fig. \ref{fig:lstm}. We have added a dropout layer after the LSTM block in Fig. \ref{fig:lstm} to avoid overfitting in the network.

\subsection{Gated Recurrent Unit (GRU)}

A GRU \cite{cho2014learning} is constructed exactly like an LSTM network, except for the output gate. A comparable GRU typically has fewer trainable parameters because it lacks the output gates. It also converges faster for the same reason. GRU has comparable performance to LSTM for a majority of sequence modeling tasks and sometimes outperform LSTM for less repetitive sequences. The general architecture of a GRU classifier is same as that of an LSTM classifier except it has a GRU block in place of the LSTM block.

\section{Result \& Validation}\label{sec:res}

In this section we validate our model by predicting attacker behavior. As mentioned in section \ref{sec:meth}, we predict the $4$ features event type ($4$ class labels), shell command with parameter ($300+$ class labels), shell command ($40+$ class labels) and shell command type ($7$ class labels) for the next event in a session. We call this set of targets the `attacker behavior'. We compare the accuracy and performance of $3$ neural networks -- TCN, LSTM, and GRU in this section. The design, simulation and testing of the neural networks are done in Matlab \cite{matlab2021mathworks}.

As listed in subsection \ref{ss:datastat}, for this simulation, we have collected a robust dataset of $3567$ `cowrie' events. These events belong to $393$ `cowrie' sessions and span over a duration of $1$ month from $4$ April 2020 to $8$ May 2020. We have collected the dataset from $5$ different `cowrie` honeypots placed all over the world. The locations of the honeypots are -- Amsterdam, Bangalore, London, Singapore, and Toronto.

To optimize the parameters of the neural networks, we have used a simple grid search. The results of the grid search for TCN are shown in subsection \ref{ss:hyper}. All the parameters are listed in \ref{app:hyper}.

\begin{table}[!ht]
    \centering
    \caption{TCN v LSTM v GRU. \\ \# training sequences = $315$; \# test sequences =  $78$.}
    \begin{tabular}{|l||c|c|c|c|c|c|}
        \hline
        & \multicolumn{3}{c|}{Accuracy (\%)} \\
        \hline
        Target & TCN & LSTM & GRU \\ \hline \hline
        Event type & $\mathbf{98.45}$ & $96.26$ & $96.51$  \\ \hline
        Command $+$ paramter & $\mathbf{85.58}$ & $80.55$ & $81.79$ \\ \hline
        Command & $\mathbf{94.64}$ & $88.87$ & $87.75$ \\ \hline
        Command type & $\mathbf{96.37}$ & $91.63$ & $90.93$ \\ \hline
    \end{tabular}
    \label{tb_res}
\end{table}

For the first test, we have randomly split the $393$ sessions in a ratio of $80:20$ into training and test subsets. Note that, each session is considered one sequence in our model and they have variable number of events in them. So, we ended up with $315$ sequences ($2839$ events) in the training dataset and $78$ ($728$ events) sequences in the test dataset. Then, we have trained our models on the training dataset. Finally, we have tested the accuracy of TCN, LSTM, and GRU in predicting the the $4$ targets for the next event in the sequence. The prediction accuracy-values are listed in Table \ref{tb_res}. It's seen that, LSTM and GRU have comparable performance, while TCN largely outperforms the other two in all $4$ cases.

While at first glance the accuracy for `command + parameter' seems low, at $~85\%$; it should be noted that this label assumes a different value for a different parameter supplied to the same command. For example \texttt{mkdir temp1} and \texttt{mkdir temp2} will be registered has different classes for this target event though they are the same command and serves the same purpose.

\begin{figure}[!h]
    \centering
    \begin{subfigure}[b]{0.48\textwidth}
        \centering
        \includegraphics[width=\textwidth]{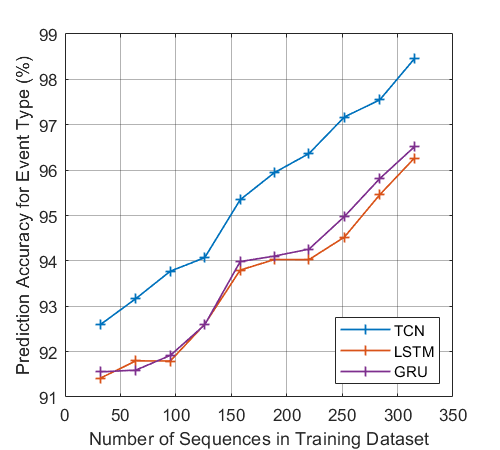}
        \caption{Target = Event type.}
        \label{fig:res11}
    \end{subfigure}
    \hfill
    \begin{subfigure}[b]{0.48\textwidth}
        \centering
        \includegraphics[width=\textwidth]{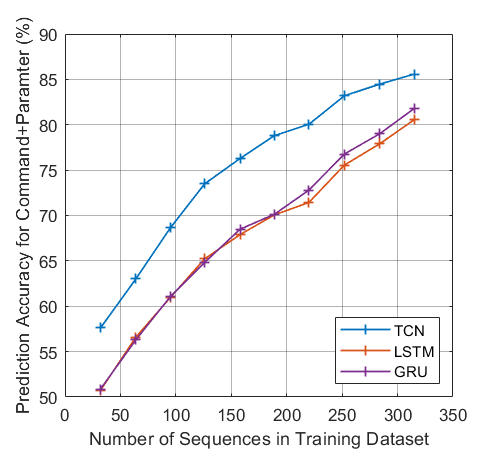}
        \caption{Target = Command + parameter.}
        \label{fig:res12}
    \end{subfigure}
    \begin{subfigure}[b]{0.48\textwidth}
        \centering
        \includegraphics[width=\textwidth]{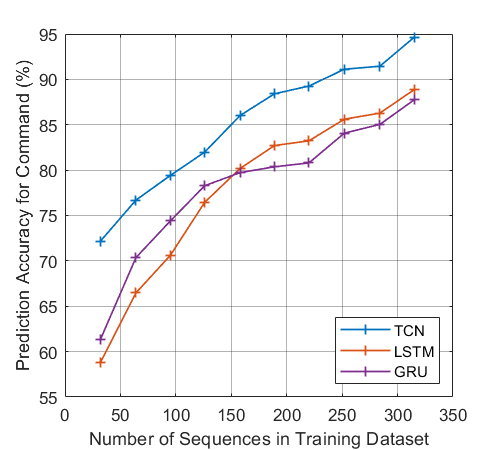}
        \caption{Target = Command.}
        \label{fig:res13}
    \end{subfigure}
    \hfill
    \begin{subfigure}[b]{0.48\textwidth}
        \centering
        \includegraphics[width=\textwidth]{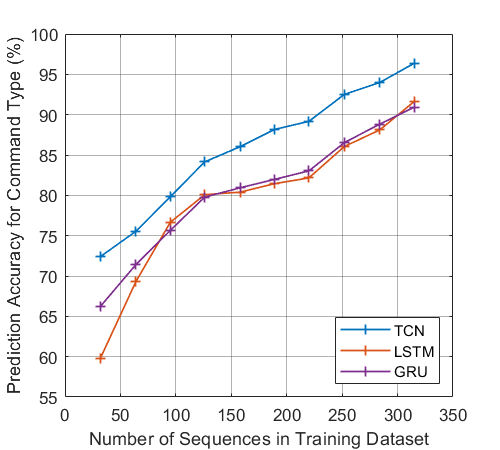}
        \caption{Target = Command Type.}
        \label{fig:res14}
    \end{subfigure}
    \caption{Accuracy vs Number of Sequences for TCN, LSTM and GRU}
    \label{fig:res1}
\end{figure}

Next, we have tested the change in accuracy with the size of the training set. We have done this for all of the four targets and the results are shown in Fig. \ref{fig:res1}. In general, the training accuracy increases with the number of training samples in all cases for all the algorithms. And just like before, LSTM and GRU perform comparably while TCN outperforms both of them by a large margin.

\begin{figure}[!h]
    \centering
    \begin{subfigure}[b]{0.23\textwidth}
        \centering
        \includegraphics[width=\textwidth]{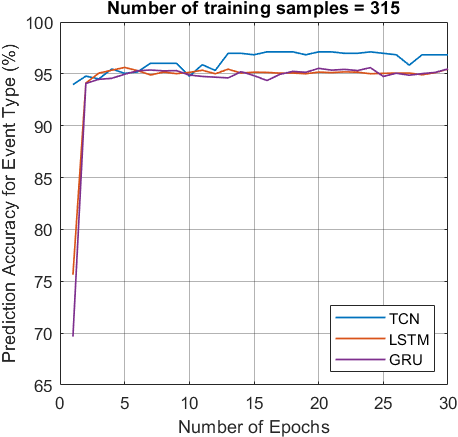}
        \caption{Event type}
        \label{fig:res21}
    \end{subfigure}
    \hfill
    \begin{subfigure}[b]{0.23\textwidth}
        \centering
        \includegraphics[width=\textwidth]{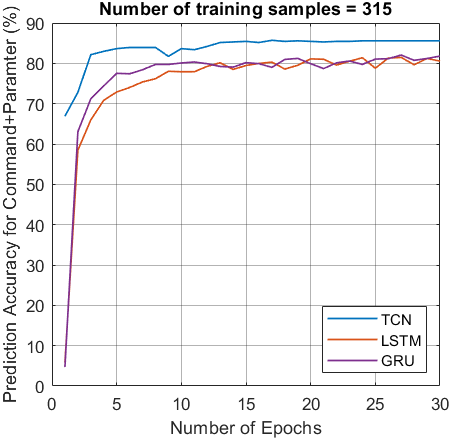}
        \caption{Command+param}
        \label{fig:res22}
    \end{subfigure}
    \begin{subfigure}[b]{0.23\textwidth}
        \centering
        \includegraphics[width=\textwidth]{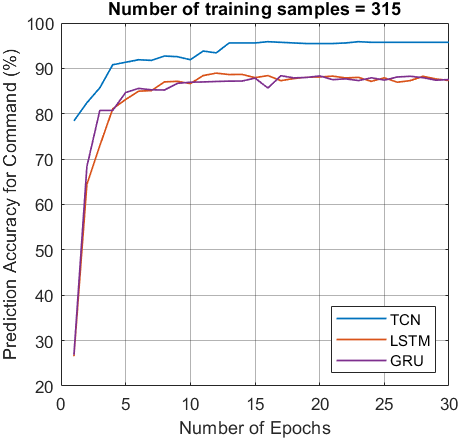}
        \caption{Command}
        \label{fig:res23}
    \end{subfigure}
    \hfill
    \begin{subfigure}[b]{0.23\textwidth}
        \centering
        \includegraphics[width=\textwidth]{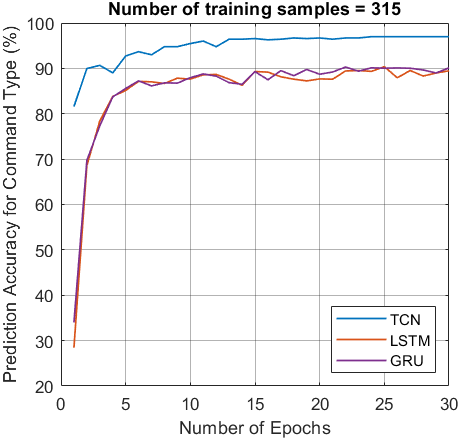}
        \caption{Command Type}
        \label{fig:res24}
    \end{subfigure}
    \caption{Accuracy vs Epochs for TCN, LSTM and GRU. \# of training sequences = $315$.}
    \label{fig:res2}
\end{figure}

\begin{figure}[!h]
    \centering
    \begin{subfigure}[b]{0.23\textwidth}
        \centering
        \includegraphics[width=\textwidth]{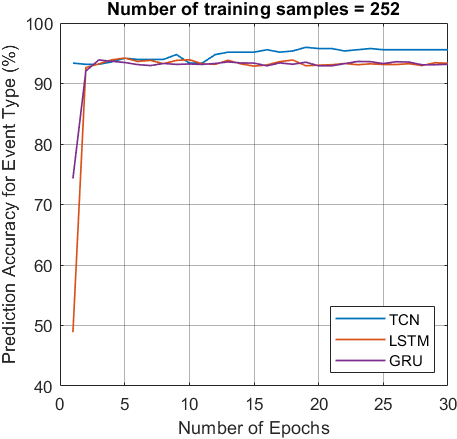}
        \caption{Event type}
        \label{fig:res31}
    \end{subfigure}
    \hfill
    \begin{subfigure}[b]{0.23\textwidth}
        \centering
        \includegraphics[width=\textwidth]{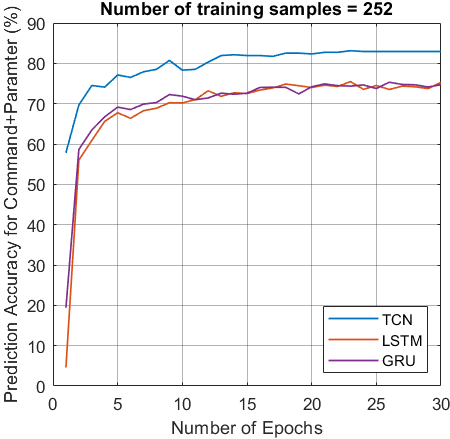}
        \caption{Command+param}
        \label{fig:res32}
    \end{subfigure}
    \begin{subfigure}[b]{0.23\textwidth}
        \centering
        \includegraphics[width=\textwidth]{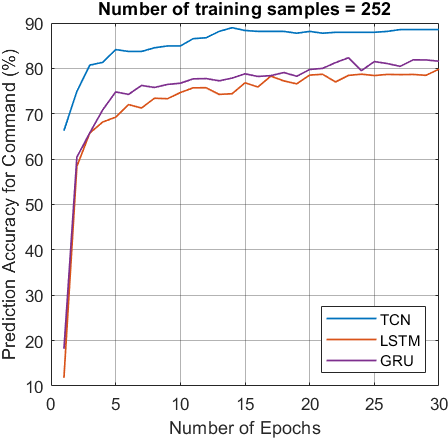}
        \caption{Command}
        \label{fig:res33}
    \end{subfigure}
    \hfill
    \begin{subfigure}[b]{0.23\textwidth}
        \centering
        \includegraphics[width=\textwidth]{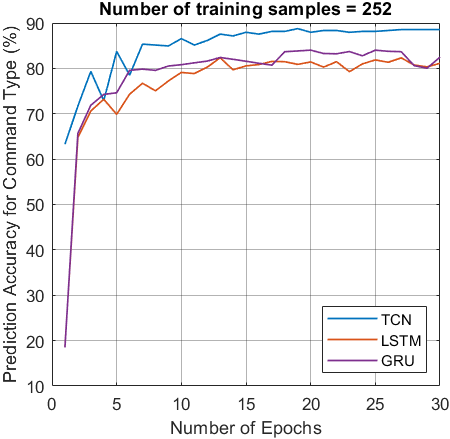}
        \caption{Command Type}
        \label{fig:res34}
    \end{subfigure}
    \caption{Accuracy vs Epochs for TCN, LSTM and GRU. \# of training sequences = $252$.}
    \label{fig:res3}
\end{figure}

Finally, we have tested how fast each algorithm converges to the final accuracy. We have again done this for all four targets. We have also compared the results for two different numbers of training sequences -- $315$ and $252$. The results are shown in figures \ref{fig:res2} and \ref{fig:res3}. These results show that TCN converges perfectly to the final accuracy well before the maximum $30$ epochs in all cases. LSTM and GRU, however, perform significantly worse than TCN.

\subsection{Optimization of TCN Parameters}\label{ss:hyper}

\begin{figure}[!ht]
    \centering
    \begin{subfigure}[b]{0.48\textwidth}
        \centering
        \includegraphics[width=\textwidth]{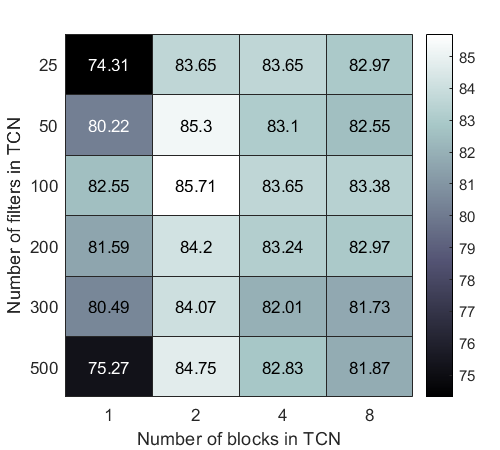}
        \caption{Heatmap of accuracy for different values of \texttt{numBlock} and \texttt{numFilt}.}
        \label{fig:res41}
    \end{subfigure}
    \hfill
    \begin{subfigure}[b]{0.48\textwidth}
        \centering
        \includegraphics[width=\textwidth]{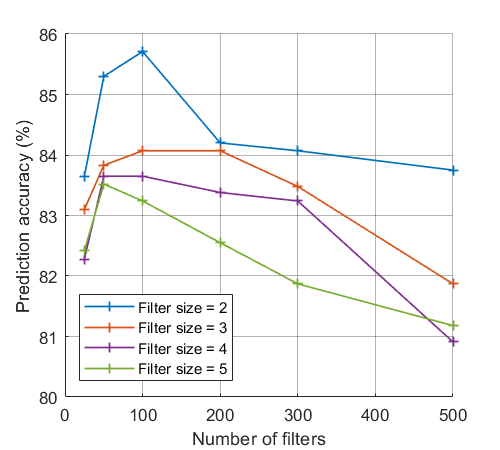}
        \caption{Accuracy vs number of filters for varying filter size.}
        \label{fig:res42}
    \end{subfigure}
\end{figure}

This subsection includes the results of the grid-search that we used to optimize the TCN parameters. Number of filters per block, number of residual blocks and filter size are the three major parameters of a TCN which in turn determine the number of learnable parameters of the network. Fig. \ref{fig:res41} shows the accuracy of our network, as a heatmap, for different combinations number of blocks and filter size. As evident from Fig. \ref{fig:res41} the model performs best for $2$ residual blocks with $100$ filters in each residual block. Furthermore, we plotted the accuracy of the network for varying filter sizes in Fig. \ref{fig:res42}. The Fig. \ref{fig:res42} shows that filter size $2$ is the optimal choice for this problem irrespective of the number of filters in each residual block.

\section{Conclusion and Future Work}

In this work we have modeled attacker behavior in IoT botnets. To model attacker behavior we have used heterogeneous threat data including shell commands input by attackers into the shell, network traffic and downloaded files. Our model incorporates the sequence of events in the attack along with the commands input into the shell. It also handles the arbitrary length of a sequence of events across various attack chains.

In this research we have also outlined a robust framework for automated analysis of the attacker behavior. To do that we have utilized CYBEX infrastructure as a service. CYBEX allows us to seamlessly automate the entire process from data collection to data analysis making our system suitable for real time implementation. Using CYBEX we have collected a robust dataset from $5$ honeypots across the world.

Finally, we have incorporated temporal convolutional networks (TCN) to predict attacker behavior. A prediction accuracy of $85-97\%$ proves the validity of our approach. We have also compared TCN  with long short term memory (LSTM) and Gated Recurrent Unit (GRU) and showed that TCN outperforms the other two by a large margin.

\appendix

\section{Parameters of Models}\label{app:hyper}.

\begin{table}[!ht]
    \label{tbl:hyper_tcn}
    \centering
    \begin{tabular}{c c}
        \hline

        \multicolumn{2}{c}{TCN Parameters} \\ \hline
        Number of residual blocks & $2$ \\
        Number of filters in each residual block & $100$ \\
        Filter size & $2$ \\
        Dropout factor & $0.02$ \\
        Maximum epochs & $30$ \\
        Minibatch size & $1$ \\
        Initial learn-rate & $0.001$ \\
        Learn-rate drop factor & $0.1$ \\
        Learn-rate drop period & $12$ epochs \\
        Gradient threshold & $1$ \\ \hline

    \end{tabular}
\end{table}

\begin{table}[!ht]
    \label{tbl:hyper_lstm}
    \centering
    \begin{tabular}{c c}
        \hline

        \multicolumn{2}{c}{LSTM Parameters} \\ \hline
        Number of hidden units & $600$ \\
        Dropout factor & $0.05$ \\
        Maximum epochs & $30$ \\
        Minibatch size & $1$ \\
        Initial learn-rate & $0.001$ \\
        Learn-rate drop factor & $0.1$ \\
        Learn-rate drop period & $12$ epochs \\
        Gradient threshold & $1$ \\ \hline

    \end{tabular}
\end{table}

\begin{table}[!ht]
    \label{tbl:hyper_gru}
    \centering
    \begin{tabular}{c c}
        \hline

        \multicolumn{2}{c}{GRU Parameters} \\ \hline
        Number of hidden units & $600$ \\
        Dropout factor & $0.05$ \\
        Maximum epochs & $30$ \\
        Minibatch size & $1$ \\
        Initial learn-rate & $0.001$ \\
        Learn-rate drop factor & $0.1$ \\
        Learn-rate drop period & $12$ epochs \\
        Gradient threshold & $1$ \\ \hline

    \end{tabular}
\end{table}

\bibliography{refs}

\begin{thebibliography}{10}
\expandafter\ifx\csname url\endcsname\relax
  \def\url#1{\texttt{#1}}\fi
\expandafter\ifx\csname urlprefix\endcsname\relax\def\urlprefix{URL }\fi
\expandafter\ifx\csname href\endcsname\relax
  \def\href#1#2{#2} \def\path#1{#1}\fi

\bibitem{sadique2021analysis}
F.~Sadique, S.~Sengupta, Analysis of attacker behavior in compromised hosts
  during command and control, in: 2021 IEEE international conference on
  communications (ICC), IEEE, 2021, p. to appear.

\bibitem{dunham2008malicious}
K.~Dunham, J.~Melnick, Malicious bots: an inside look into the cyber-criminal
  underground of the internet, CrC Press, 2008.

\bibitem{antonakakis2017understanding}
M.~Antonakakis, T.~April, M.~Bailey, M.~Bernhard, E.~Bursztein, J.~Cochran,
  Z.~Durumeric, J.~A. Halderman, L.~Invernizzi, M.~Kallitsis, et~al.,
  Understanding the mirai botnet, in: 26th $\{$USENIX$\}$ security symposium
  ($\{$USENIX$\}$ Security 17), 2017, pp. 1093--1110.

\bibitem{stone2009your}
B.~Stone-Gross, M.~Cova, L.~Cavallaro, B.~Gilbert, M.~Szydlowski, R.~Kemmerer,
  C.~Kruegel, G.~Vigna, Your botnet is my botnet: analysis of a botnet
  takeover, in: Proceedings of the 16th ACM conference on Computer and
  communications security, 2009, pp. 635--647.

\bibitem{shin2010conficker}
S.~Shin, G.~Gu, Conficker and beyond: a large-scale empirical study, in:
  Proceedings of the 26th Annual Computer Security Applications Conference,
  2010, pp. 151--160.

\bibitem{puri2003bots}
R.~Puri, Bots \& botnet: An overview, SANS Institute 3 (2003).

\bibitem{feily2009survey}
M.~Feily, A.~Shahrestani, S.~Ramadass, A survey of botnet and botnet detection,
  in: 2009 3rd International Conference on Emerging Security Information,
  Systems and Technologies, IEEE, 2009.

\bibitem{kolias2017ddos}
C.~Kolias, G.~Kambourakis, A.~Stavrou, J.~Voas, Ddos in the iot: Mirai and
  other botnets, Computer 50~(7) (2017) 80--84.

\bibitem{liao2013intrusion}
H.-J. Liao, C.-H.~R. Lin, Y.-C. Lin, K.-Y. Tung, Intrusion detection system: A
  comprehensive review, Journal of Network and Computer Applications 36~(1)
  (2013) 16--24.

\bibitem{karim2014botnet}
A.~Karim, R.~B. Salleh, M.~Shiraz, S.~A.~A. Shah, I.~Awan, N.~B. Anuar, Botnet
  detection techniques: review, future trends, and issues, Journal of Zhejiang
  University SCIENCE C 15~(11) (2014) 943--983.

\bibitem{karasaridis2007wide}
A.~Karasaridis, B.~Rexroad, D.~A. Hoeflin, et~al., Wide-scale botnet detection
  and characterization., HotBots 7 (2007) 7--7.

\bibitem{binkley2006algorithm}
J.~R. Binkley, S.~Singh, An algorithm for anomaly-based botnet detection.,
  SRUTI 6 (2006) 7--7.

\bibitem{gu2008botsniffer}
G.~Gu, J.~Zhang, W.~Lee, Botsniffer: Detecting botnet command and control
  channels in network traffic (2008).

\bibitem{choi2007botnet}
H.~Choi, H.~Lee, H.~Lee, H.~Kim, Botnet detection by monitoring group
  activities in dns traffic, in: 7th IEEE International Conference on Computer
  and Information Technology (CIT 2007), IEEE, 2007.

\bibitem{villamarin2008identifying}
R.~Villamar{\'\i}n-Salom{\'o}n, J.~C. Brustoloni, Identifying botnets using
  anomaly detection techniques applied to dns traffic, in: 2008 5th IEEE
  Consumer Communications and Networking Conference, IEEE, 2008.

\bibitem{dagon2005botnet}
D.~Dagon, Botnet detection and response, in: OARC workshop, 2005.

\bibitem{schonewille2006domain}
A.~Schonewille, D.-J. Van~Helmond, The domain name service as an ids, Research
  Project for the Master System-and Network Engineering at the University of
  Amsterdam (2006).

\bibitem{murugan2011system}
S.~Murugan, K.~Kuppusamy, System and methodology for unknown malware attack
  (2011).

\bibitem{creech2013semantic}
G.~Creech, J.~Hu, A semantic approach to host-based intrusion detection systems
  using contiguousand discontiguous system call patterns, IEEE Transactions on
  Computers 63~(4) (2013) 807--819.

\bibitem{ge2012effective}
L.~Ge, H.~Liu, D.~Zhang, W.~Yu, R.~Hardy, R.~Reschly, On effective sampling
  techniques for host-based intrusion detection in manet, in: MILCOM 2012-2012
  IEEE Military Communications Conference, IEEE, 2012, pp. 1--6.

\bibitem{sadique2021cybersecurity}
F.~Sadique, I.~Astaburuaga, R.~Kaul, S.~Sengupta, S.~Badsha, J.~Schnebly,
  A.~Cassell, J.~Springer, N.~Latourrette, S.~M. Dascalu, Cybersecurity
  information exchange with privacy (cybex-p) and tahoe -- a cyberthreat
  language (2021).
\newblock \href {http://arxiv.org/abs/2106.01632} {\path{arXiv:2106.01632}}.

\bibitem{sadique2019system}
F.~Sadique, K.~Bakhshaliyev, J.~Springer, S.~Sengupta, A system architecture of
  cybersecurity information exchange with privacy (cybex-p), in: 2019 IEEE 9th
  Annual Computing and Communication Workshop and Conference (CCWC), IEEE,
  2019, pp. 0493--0498.

\bibitem{bai2018empirical}
S.~Bai, J.~Z. Kolter, V.~Koltun, An empirical evaluation of generic
  convolutional and recurrent networks for sequence modeling, arXiv preprint
  arXiv:1803.01271 (2018).

\bibitem{shrivastava2019attack}
R.~K. Shrivastava, B.~Bashir, C.~Hota, Attack detection and forensics using
  honeypot in iot environment, in: International Conference on Distributed
  Computing and Internet Technology, Springer, 2019.

\bibitem{oosterhof2016cowrie}
M.~Oosterhof, Cowrie ssh/telnet honeypot (2016).

\bibitem{rade2018temporal}
R.~Rade, S.~Deshmukh, R.~Nene, A.~S. Wadekar, A.~Unny, Temporal and stochastic
  modelling of attacker behaviour, in: International Conference on Intelligent
  Information Technologies, Springer, 2018.

\bibitem{deshmukh2019attacker}
S.~Deshmukh, R.~Rade, D.~Kazi, et~al., Attacker behaviour profiling using
  stochastic ensemble of hidden markov models, arXiv preprint arXiv:1905.11824
  (2019).

\bibitem{spitzner2003honeynet}
L.~Spitzner, The honeynet project: Trapping the hackers, IEEE Security \&
  Privacy 1~(2) (2003) 15--23.

\bibitem{dierks2008transport}
T.~Dierks, E.~Rescorla, The transport layer security (tls) protocol version 1.2
  (2008).

\bibitem{goodfellow2016deep}
I.~Goodfellow, Y.~Bengio, A.~Courville, Y.~Bengio, Deep learning, Vol.~1, MIT
  press Cambridge, 2016.

\bibitem{hochreiter1997long}
S.~Hochreiter, J.~Schmidhuber, Long short-term memory, Neural computation 9~(8)
  (1997) 1735--1780.

\bibitem{mikolov2011extensions}
T.~Mikolov, S.~Kombrink, L.~Burget, J.~{\v{C}}ernock{\`y}, S.~Khudanpur,
  Extensions of recurrent neural network language model, in: 2011 IEEE
  international conference on acoustics, speech and signal processing (ICASSP),
  IEEE, 2011, pp. 5528--5531.

\bibitem{cho2014learning}
K.~Cho, B.~van Merrienboer, C.~Gulcehre, D.~Bahdanau, F.~Bougares, H.~Schwenk,
  Y.~Bengio, Learning phrase representations using rnn encoder-decoder for
  statistical machine translation (2014).
\newblock \href {http://arxiv.org/abs/1406.1078} {\path{arXiv:1406.1078}}.

\bibitem{matlab2021mathworks}
MATLAB, version 9.10.0.1602886 (R2021a), The MathWorks Inc., Natick,
  Massachusetts, 2021.

\end{thebibliography}

\end{document}